\documentclass[useAMS,usenatbib]{mn2e}

\usepackage{times}
\usepackage{multirow}
\usepackage[T1]{fontenc} 
\usepackage{aecompl}

\usepackage{psfrag}
\usepackage{pspicture}
\usepackage{graphics}
\usepackage{graphicx}
\usepackage{amsmath}
\usepackage{color}

\title[H$\alpha$ survey of merging clusters]{The role of cluster mergers and travelling shocks in shaping the H$\alpha$ luminosity function at $\bf z\sim0.2$: `sausage' and `toothbrush' clusters}
\author[A. Stroe et al.]{Andra Stroe$^{1}$\thanks{E-mail: astroe@strw.leidenuniv.nl}, David Sobral$^{1}$\thanks{VENI Fellow}, Huub J. A. R\"ottgering$^{1}$, Reinout J. van Weeren$^{2}$\thanks{NASA Einstein Postdoctoral Fellow}\\
$^{1}$Leiden Observatory, Leiden University, P.O.\ Box 9513, NL-2300 RA Leiden, The Netherlands\\
$^{2}$Harvard Smithsonian Center for Astrophysics (CfA - SAO), 60 Garden Street Cambridge, MA 02138\\
}

\begin{document}
\date{Accepted 2013 November 22.  Received 2013 November 18; in original form 2013 September 11}
\maketitle

\begin{abstract}
The most extreme cluster mergers can lead to massive cluster-wide travelling shock waves. The CIZA J2242.8+5301 (`sausage') and 1RXS J0603.3+4213 (`toothbrush') clusters ($z\sim0.2$) host enormous radio-emitting shocks with simple geometry. We investigate the role of mergers and shocks in shaping the H$\alpha$ luminosity function, using custom-made narrow-band filters matching the cluster redshifts mounted on the INT. We surveyed $\sim0.28$~deg$^2$ for each cluster and found $181$ line emitters in the `sausage' (volume of $3.371\times10^3$~Mpc$^3$ for H$\alpha$ at $z=0.1945$) and $141$ in the `toothbrush' ($4.546\times10^3$~Mpc$^3$ for H$\alpha$ at $z=0.225$), out of which $49$ (`sausage') and $30$ (`toothbrush') are expected to be H$\alpha$. We build luminosity functions for the field-of-view down to an average limiting star formation rate of $0.14$ M$_{\sun}$ yr$^{-1}$, find good agreement with field luminosity functions at $z=0.2$, but significant differences between the shapes of the luminosity functions for the two clusters. We discover extended, tens-of-kpc-wide H$\alpha$ haloes in galaxies neighbouring relics, which were possibly disrupted by the passage of the shock wave. By comparing the `sausage' cluster with blank fields and other clusters, we also uncover an order of magnitude boost (at $9\sigma$ level) in the normalisation $\phi^*$ of the luminosity function in the relic areas. Our results suggest that cluster mergers may play an important role in the evolution of cluster galaxies through shock-induced star formation.
\end{abstract}


\section{INTRODUCTION}\label{sec:intro}

Tracing star formation across cosmic time and in different environments is instrumental to our understanding of galaxy evolution. Studies have shown that star formation (SF) activity has steadily declined since redshift $z\sim2-3$ \citep{1996ApJ...460L...1L, 2006ApJ...651..142H, 2013MNRAS.428.1128S}. There is also a striking difference between field galaxies and systems within cluster environments: the fraction of blue, actively star forming, late-type galaxies is systematically lower in the latter \citep{1980ApJ...236..351D, 2003MNRAS.346..601G}. Multiple authors \citep[e.g.][]{2002MNRAS.334..673L, 2003ApJ...584..210G, 2004AJ....128.2677T, 2004MNRAS.348.1355B} have therefore concluded that the SF rate is strongly dependent on local galaxy density. Early-type galaxies, predominately found in cluster environments, are more massive than their star-forming counterparts \citep{2003MNRAS.341...33K}. However, \citet{2010ApJ...721..193P} have shown that the effect of mass can be distinguished from that of environment, although both effects have important contributions. While the shape is similar, the overall normalisation of the H$\alpha$ luminosity function (LF) in low-redshift $z\sim0.2-0.4$ rich clusters is $\sim50$ per cent lower than for the field \citep{2002MNRAS.335...10B, 2001ApJ...549..820C, 2004MNRAS.354.1103K}. The difference between cluster and field spiral galaxies is evident also from radio observations of neutral hydrogen: cluster spirals contain significantly less HI gas than their field counterparts \citep[e.g.][]{1990AJ....100..604C}. There has been evidence of ram pressure stripping of the HI and H$\alpha$ gas in infalling cluster galaxies, such as tidal tails and filaments \citep{2001ApJ...563L..23G,2005A&A...437L..19O}. Interaction of galaxies with the intra-cluster medium is evident within high-redshift galaxies. Alignment of optical and radio continuum in active galaxies \citep[e.g.][]{1987AJ.....93.1307D,1987ApJ...321L..29M} has prompted an interpretation where the travelling shock front at the wake of the radio lobes triggers star formation \citep{1989MNRAS.239P...1R}. While the shock would evacuate the radio lobe of its hot gas and shock-heat the surrounding medium, the cool clouds within the radio lobes would collapse to form stars in this over-pressured environment.

The luminous components in galaxy clusters can reveal the complex evolution of baryonic matter through cosmic time and the interplay between the intra-cluster medium (ICM) and the cluster galaxies. Driven by mergers with other galaxy clusters and groups, the growth of clusters releases copious amounts of gravitational energy deposited into the ICM \citep[e.g.][]{2002ASSL..272....1S}, with direct evidence from the X-ray and radio in the form of shock fronts. The strongest cluster major mergers can produce cluster-wide shock waves that travel through the intracluster medium and accelerate particles through the diffusive shock acceleration mechanism \citep{1983RPPh...46..973D}. Shock waves can be detected in the X-ray as density/temperature discontinuities \citep[e.g.][]{2005ApJ...627..733M} or in the radio bands as relics: elongated, diffuse synchrotron emitting areas located at the periphery of merging clusters \citep[see review paper by][]{2012A&ARv..20...54F}.

The precise effect of the merging history of a cluster on the evolution of galaxies is unknown. The profound impact of the travelling shock waves on the ICM raises questions regarding the interaction between the galaxies within the cluster and the shock front, coupled with the merging nature of the cluster. What is the effect of the travelling shock wave on the ISM of galaxies and their star formation activity? Are there any morphological differences between galaxies in the pre- and post-shock regions? 

To measure the effect of the travelling shock waves on the SF activity within clusters, we need to trace massive, newborn stars within the cluster galaxies. These stars emit strong, ionising UV radiation. While this is often absorbed in nearby regions, it is then re-emitted through a variety of processes, such as far-infra-red (FIR) black-body emission (from the heated dust) and recombination lines, of which the H$\alpha$ recombination line is the strongest and best calibrated. H$\alpha$ suffers only from modest dust extinction and is much more sensitive to instantaneous SF than the UV/FIR. An excellent way to detect line emitters over large areas entails using narrow-band filters tuned to be sensitive to the H$\alpha$ emission redshifted at the distance of your source of interest. Subtracting the broad-band (BB) emission from the narrow-band (NB) emission singles out sources which have line emission. In recent years, large H$\alpha$ surveys of blank fields have been carried out up to redshift $2.23$ \citep{2003ApJ...586L.115F, 2007ApJ...657..738L, 2008ApJS..175..128S, 2008ApJ...677..169V, 2008MNRAS.388.1473G, 2009MNRAS.398...75S, 2011ApJ...726..109L, 2011MNRAS.411..675S, 2013MNRAS.433..796D, 2013MNRAS.428.1128S}. Studies of clusters have been limited to low-redshift relaxed, rich clusters \citep{2001ApJ...549..820C, 2002MNRAS.335...10B, 2004MNRAS.354.1103K} or higher redshift ($z=0.8,1.47$), almost virialised clusters \citep{2010MNRAS.403.1611K, 2010MNRAS.402.1980H, 2013MNRAS.434..423K}.

Until now, it has been difficult to address the merging cluster H$\alpha$ LF, owing to a lack of suitable systems. To directly interpret observables such as the radio/X-ray morphology and galaxy distribution in terms of simple physical parameters, we would like to study equal mass systems merging in the plane of the sky with a low impact parameter, ideally at a moment in time when shocks are prominently present. Recently, \citet{2010Sci...330..347V} and \citet{2012A&A...546A.124V} have discovered spectacular Mpc-size, coherent radio shocks in two merging clusters with the required properties. CIZA J2242.8+5301 (nicknamed the `sausage') is a binary merging cluster at $z=0.192$ (see Figure~\ref{fig:image}, left). A $2$ Mpc relic at the northern periphery dwarfs a smaller one located symmetrically across the cluster centre. The northern relic is marked by signatures of cooling, synchrotron-emitting particles in the post-shock region. \citet{2013A&A...555A.110S} discovered a bi-modality in the orientation of the head-tail radio galaxies which act as a tracer of the two merging clusters. Cluster 1RXS J0603.3+4213 (nicknamed the `toothbrush'), at $z=0.225$ (see Figure~\ref{fig:image}, right), hosts a $2.5$ Mpc long and straight radio relic and also a smaller relic towards the south-east. Both clusters have a disturbed, elongated X-ray morphology \citep{2013MNRAS.429.2617O, 2013MNRAS.433..812O, 2013PASJ...65...16A}, indicating they are post-core passage mergers. The different merger histories and post-merger timescales of the two clusters might provide valuable information about the evolution of the interaction between the galaxies and the shock front.

In this paper, we characterise the nature of galaxies in highly disturbed, $z\sim0.2$, merging clusters hosting radio relics. We explore the imprint of the travelling shock wave on the morphology of the ionised gas within galaxies and on the H$\alpha$ luminosity function. The paper is structured in the following way: in \S\ref{sec:obs-reduction} we give an overview of the observations and the data reduction, in \S\ref{sec:results} we present the results, \S\ref{sec:discussion} shows the implications for galaxy evolution within merging clusters. The main points are summarised in \S\ref{sec:conclusion}. A flat, $\Lambda$CDM cosmology with $H_{0}=70.5$~km~s$^{-1}$~Mpc$^{-1}$, matter density $\Omega_M=0.27$ and dark energy density $\Omega_{\Lambda}=0.73$ is assumed \citep{2009ApJS..180..306D}. At the redshift of the `sausage' cluster, $1$~arcmin corresponds to $0.191$~Mpc, while for the `toothbrush' it measures $0.216$~Mpc. All images are in the J2000 coordinate system. We use the online cosmological calculator described in \citet{2006PASP..118.1711W}. All magnitudes are in the AB system, except where noted otherwise. 

\section{OBSERVATIONS \& DATA REDUCTION}
\label{sec:obs-reduction}
\subsection{Isaac Newton Telescope observations}
\label{sec:obs-reduction:obs}

\begin{figure}
\begin{center}
\includegraphics[trim =0cm 0cm 0cm 0cm, width=0.46\textwidth]{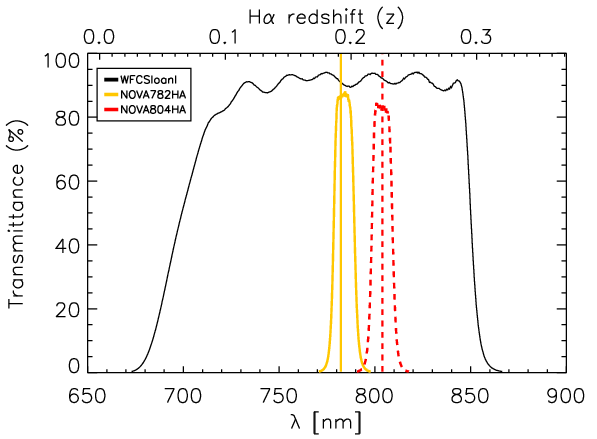}
\end{center}
\caption{Transmittance profiles for the three filters used in this analysis. The solid black line marks the profile of the BB filter. The solid yellow line defines filter NOVA782HA and the dashed red line the filter NOVA804HA. The top x-axis marks the redshifted H$\alpha$ coverage of the filters. The vertical lines mark the redshifts of the two clusters.}
\label{fig:transmittance}
\end{figure}

Optical imaging data were obtained with the Wide Field Camera (WFC)\footnote{http://www.ing.iac.es/engineering/detectors/ultra\_wfc.htm} installed at the prime focus of the Isaac Newton Telescope (INT) \footnote{http://www.ing.iac.es/Astronomy/telescopes/int/}. The large field of view (FOV) enables us to instantaneously capture each cluster and its outskirts within $0.3$ deg$^2$ using a single method (equivalent to the area shown in the radio intensity in Fig.~\ref{fig:image}). This corresponds to an area of $6.5\times6.5$~Mpc, thirteen times the expected virial size of the cluster \citep{1986RvMP...58....1S}. The instrument is a mosaic of four chips of $2048\times4100$ pixels, arranged in a square with a $1'$ inter-chip spacing. The CCDs have a pixel scale of $0.33$ arcsec pixel$^{-1}$. The FOV is $34.2'\times34.2'$, with the top north-western corner missing (e.g. Fig.~\ref{fig:image}). We used the WFCSloanI broad band (BB) filter centred at $7743$\,{\AA} and full width at half maximum (FWHM) of $1519$ {\AA}, and custom made narrow band (NB) filters NOVA782HA ($\lambda_\mathrm{c}=7839$\;{\AA}) and NOVA804HA ($\lambda_\mathrm{c}=8038.5$\;{\AA}), both $110$ {\AA} wide at FWHM. The NB filters were designed to match the redshifted H$\alpha$ (restframe $\lambda=6562.8$\;{\AA}) emission at the redshift of the clusters ($z=0.1945$ and $z=0.2250$, respectively). Figure~\ref{fig:transmittance} presents the filter response and the redshift of H$\alpha$ emission they trace. The values are also summarised in Table~\ref{tab:filters}.

The two fields were observed between October 13--22, 2012. Conditions were photometric for six nights, when the seeing varied between $0.9$ and $1.1$ arcsec. Individual exposures of $200$s in the BB and $600$s in the NB were taken in 5 jittered positions to cover the chip gaps and obtain a contiguous coverage of the FOV. 

\begin{table}
\begin{center}
\caption{Filter properties: type (narrow band, NB, or broad band, BB), central wavelength, full width at half maximum and the redshift range $z_{\mathrm{H}\alpha}$ for which the H$\alpha$ line is detected within the FWHM of the narrow band filters.}
\begin{tabular}{l l l l l}
\hline
\hline
Filter & Type & $\lambda_\mathrm{c}$ ({\AA}) & FWHM ({\AA}) & $z_{\mathrm{H}\alpha}$ \\
\hline
NOVA782HA & NB & $7839.0$   & $110$ & $0.1865-0.2025$ \\
NOVA804HA & NB & $8038.5$ & $110$ & $0.2170-0.2330$ \\
WFCSloanI & BB & $7743.0$   & $1519$ & -- \\
\hline
\end{tabular}
\label{tab:filters}
\end{center}
\end{table}

Both clusters are found at low Galactic latitudes ($-5.1^{\circ}$ for the `sausage' and $9.7^{\circ}$ for the `toothbrush') and suffer from substantial dust extinction: $0.76$ mag and $0.38$ mag at the NB filter wavelength, respectively (see Fig.~\ref{fig:dust}). The total integration times were driven by the different extinction values for the two clusters to reach a similar flux limit and obtain comparable results for the two targets. The details of the observations can be found in Table~\ref{tab:observations}. 

\subsection{Reduction}
\label{sec:obs-reduction:red}

We created a pipeline following the standard procedure for optical data reduction. Before reducing the data, bad frames were rejected, such as frames affected by significant cloud extinction ($>0.3$ mag), drifting pointing, poor focussing and read-out issues. This resulted in removal of $\sim40$ per cent of the frames (see Table~\ref{tab:observations}). The data were then split into individual frames for each camera chip and processed independently.

\begin{table*}
\begin{center}
\caption{Details of the observations: total integration times, effective integration times after removing bad frames and observing dates.}
\begin{tabular}{l l c c c c l}
\hline
\hline
Field & RA & DEC & Filter & Int. Time (ks) & Eff. Time (ks) & Dates \\
\hline
\multirow{2}{*}{`Sausage'} & \multirow{2}{*}{$22^{h}42^{m}50^{s}$} & \multirow{2}{*}{$53^{\circ}06'30''$} & NOVA782HA & $66.9$ & $35.4$ & October 13--15, 20--22, 2012 \\
 & & & WFCSloanI & \phantom{0}$7.4$ & \phantom{0}$3.2$ & October 15, 20--22, 2012  \\ \hline
\multirow{2}{*}{`Toothbrush'} & \multirow{2}{*}{$06^{h}03^{m}30^{s}$} & \multirow{2}{*}{$42^{\circ}17'30''$} & NOVA804HA & $37.8$ & $24.0$ & October 13--16, 20, 21, 2012 \\
 & & &  WFCSloanI & \phantom{0}$5.0$ & \phantom{0}$5.2$ & October 15, 16, 21, 2012 \\
\hline
\end{tabular}
\label{tab:observations}
\end{center}
\end{table*}

A `master flat' was obtained for each filter by median-combining and normalising all of the available sky flats from all the nights. One `master bias' was obtained for each night. After the science frames were bias-subtracted and flattened, astrometric solutions with $0.3-0.4$ arcsec root-mean-square (rms) were obtained with {\sc SCAMP}  \citep{2006ASPC..351..112B} and the USNO-B1.0 catalogue red magnitude Imag \citep{2003AJ....125..984M}.

The background noise level was not constant after flat-fielding, but presented regular, large-scale patterns not associated with real sky features. This effect, called fringing, is caused by thin-film interference in the CCD and specifically affects the red part of the optical spectrum, where our data were taken \citep{2000ASPC..216..415L}. It is crucial to remove this additive contribution from the science frames to robustly extract even the lowest signal-to-noise (S/N) sources. The ripples across the image are tied to positions in the sky and cannot be simply removed with normal flattening. Therefore, we produced a `super-flat' by median-combining the science frames with sources masked. We used this `super-flat' to self-flatten the data, which eliminated the fringing. 

In order to mask non-responsive or hot pixels, we blanked pixels that deviated by more than $3\sigma$ from the local median value of the flat. 

\subsection{Photometry}
\label{sec:obs-reduction:photometry}
Each frame was photometrically calibrated before co-adding using the USNO-B1.0 catalogue. Typically 200-300 sources were matched to the USNO-B1.0 Imag catalogue \citep{2003AJ....125..984M} and the differences between the inferred magnitudes and the ones from the USNO-B1.0 were computed. We used the median of these differences as our initial zero-point (ZP), compared that with ZP=25, and derived the appropriate scaling factor to set the magnitude ZP of each image to 25. We note that the individual $0.2$ mag uncertainties of the USNO-B1.0 Imag catalogue naturally lead to a scatter in the magnitude distribution of about $0.2$ mag, but that we always match a sufficiently large number ($\sim250$) of sources for the median to provide a robust ZP (error $\sim0.05$ mag) measurement. This step ensures the BB and NB frames are on the same magnitude scale. We finally combined all the calibrated frames, normalised to the same ZP, for each field and filter using {\sc SWarp}  \citep{2002ASPC..281..228B}. This included a mesh-based sky background  subtraction.

To check the consistency of these magnitudes with the Sloan AB system, we also performed an extra two-step calibration. The Sloan survey does not cover our targets, but the USNO-B1.0 catalogue \citep{2003AJ....125..984M} provides an all-sky coverage in three optical colours. In a first step, we found transformations between SDSS i magnitudes and USNO-B1.0 Imag magnitudes. Secondly, we tied our BB magnitude system to the USNO-B1.0 red magnitude Imag. 

We chose a $3^{\circ}\times3^{\circ}$ patch in the Galactic plane with both SDSS DR7 \citep{2009ApJS..182..543A, 2009yCat.2294....0A} and USNO-B1.0 coverage. We matched the two catalogues using a search radius of $1$ arcsec and flagged bright, saturated sources, faint sources where either catalogue became incomplete and sources which deviated by more than $3\sigma$ from the mean distribution. This resulted in a sample of $177000$ sources. We used three USNO-B1.0 optical bands and four SDSS filters (u, g, r, i) to look for any correlations and concluded that the transformation between the SDSS i and USNO-B1.0 red magnitude Imag is well described by a constant offset of $+0.54$ mag. We matched the USNO-B1.0 catalogue with the BB catalogues for our two targets and removed sources at the bright and faint end. Using $\sim7800$ matched sources for the `sausage' and $\sim4300$ for the `toothbrush', we found constant offsets between our i band magnitudes and the USNO-B1.0 Imag measurements of $-0.67$ and $-0.71$, respectively. 

We explored relationships with the red and blue USNO-B1.0 magnitudes, but found no significant trends. We concluded that the INT filters are comparable (within $5$ per cent) to the SDSS DR7 ones, therefore we used the SDSS ZP to obtain correct BB magnitudes.

\begin{figure*}
\begin{center}
\includegraphics[trim=0cm 0cm 0cm 0cm, width=0.495\textwidth]{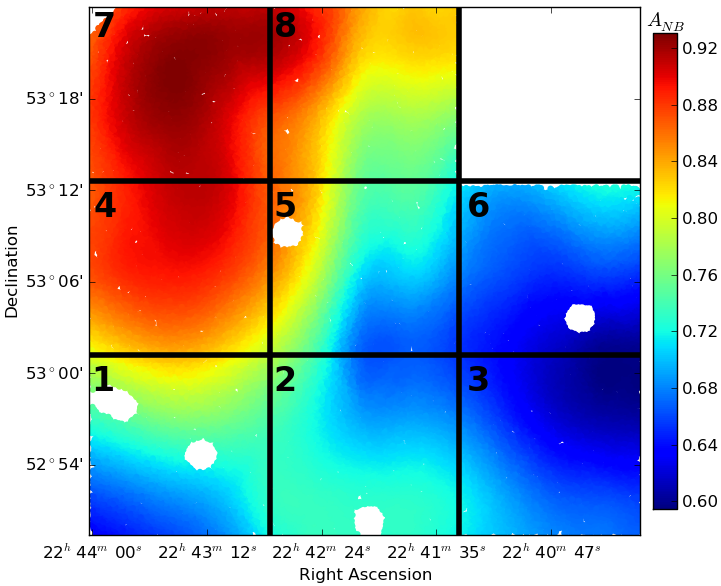}
\includegraphics[trim=0cm 0cm 0cm 0cm, width=0.495\textwidth]{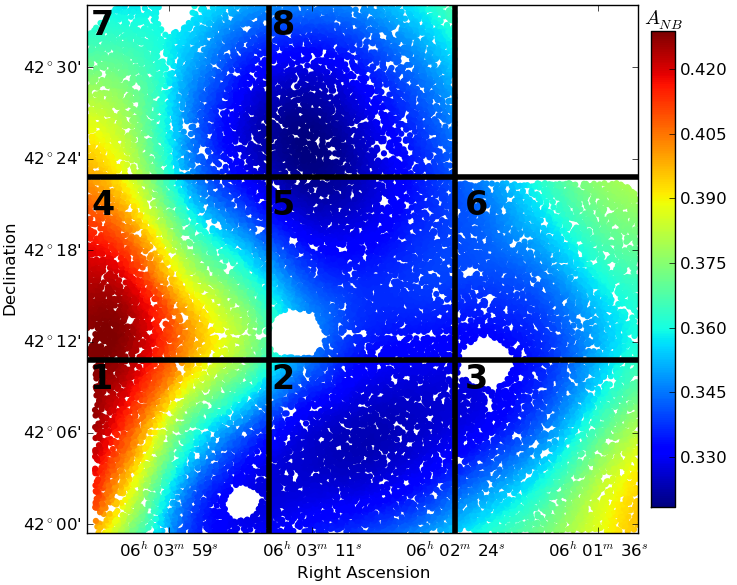}
\end{center}
\caption{Narrow band `dust screens' for the two fields in units of magnitudes. The dust extinction is predicted for wavelength $7839$\;{\AA} for the `sausage' and $8038.5$\;{\AA} for the `toothbrush'. The masked disks are the regions around bright stars that were not taken into account for the analysis. Notice the different scales of the two figures. The variations in dust extinction in the `toothbrush' field are much smaller than the `sausage'. \textit{Left}: `Sausage'. \textit{Right}: `Toothbrush'. Due to the high variability of the dust extinction, the FOV has been divided into eight areas to perform the completeness study.}
\label{fig:dust}
\end{figure*}

\subsection{Source extraction}
\label{sec:obs-reduction:extraction}
Source detection was performed using the {\sc SExtractor} package \citep{1996A&AS..117..393B}. Bright stars were masked to prevent detection of spurious sources within the `bright haloes' around them (see Fig.~\ref{fig:image} for missing coverage). Fluxes were measured in $5$ arcsec apertures corresponding to physical sizes of $\sim17$~kpc for the `sausage' and the `toothbrush'. Because of the slightly different properties of the individual chips and different exposure times, sources were detected on each individual chip independently.

We measured the rms noise level in $1000$ randomly placed apertures across the field. We repeated the experiment $100$ times for each target and each filter to minimise statistical variance. The median of the rms measurements coincided with the values reported by {\sc SExtractor} and we proceeded to use those. The average $3\sigma$ limiting magnitude (measured within $5$ arcsec apertures) for the `sausage' NB observations is $21.7$, while the BB is limited to $22.1$. The NB `toothbrush' limiting magnitude is $22.2$ and the BB goes down to magnitude $22.7$. These are observed measurements, not corrected for Galactic dust extinction (see \S\ref{sec:obs-reduction:dust}). The total number of sources detected by  {\sc SExtractor} in each filter is given in Table~\ref{tab:sources}. 

\begin{figure*}
\begin{center}
\setlength\fboxsep{0pt}
\setlength\fboxrule{3.0pt}
\fbox{\includegraphics[trim=0cm 0cm 0cm 0cm, width=0.46\textwidth]{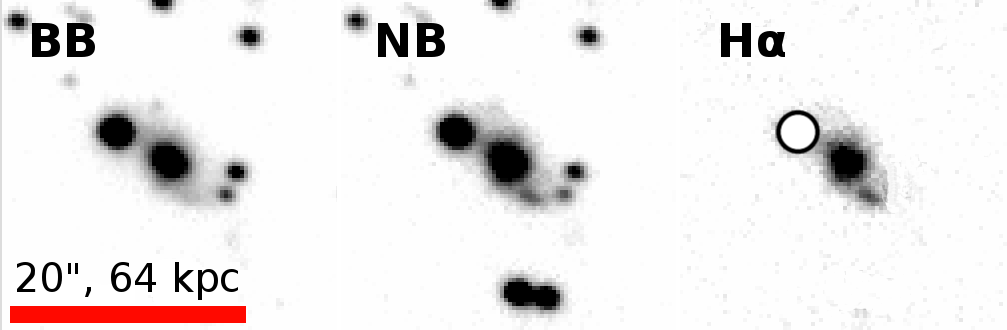}}
\fbox{\includegraphics[trim=0cm 0cm 0cm 0cm, width=0.46\textwidth]{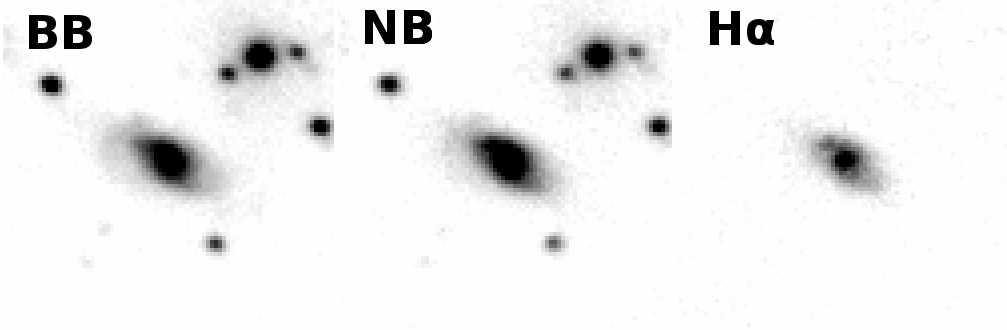}}
\fbox{\includegraphics[trim=0cm 0cm 0cm 0cm, width=0.46\textwidth]{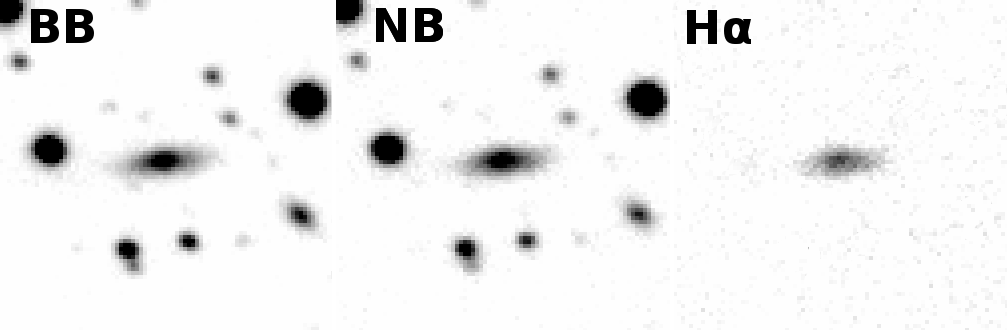}}
\fbox{\includegraphics[trim=0cm 0cm 0cm 0cm, width=0.46\textwidth]{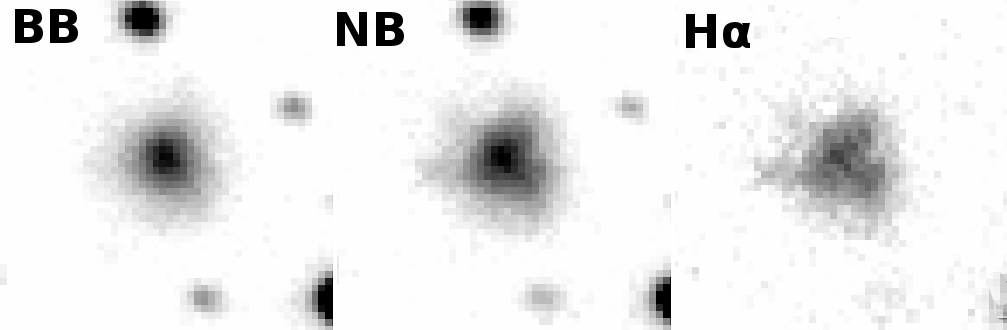}}
\fbox{\includegraphics[trim=0cm 0cm 0cm 0cm, width=0.46\textwidth]{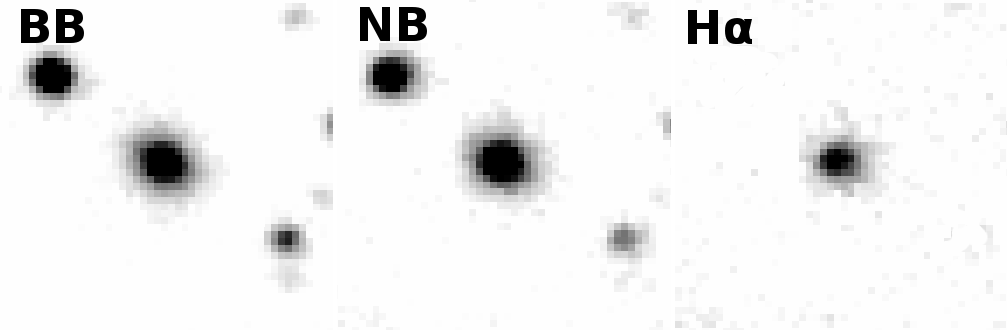}}
\fbox{\includegraphics[trim=0cm 0cm 0cm 0cm, width=0.46\textwidth]{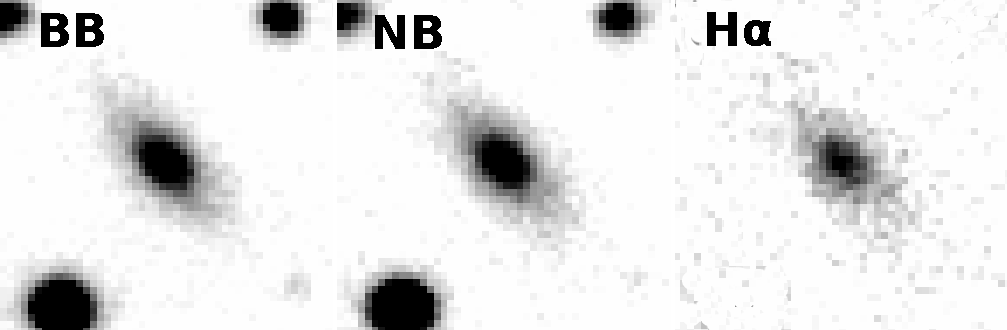}}
\fbox{\includegraphics[trim=0cm 0cm 0cm 0cm, width=0.46\textwidth]{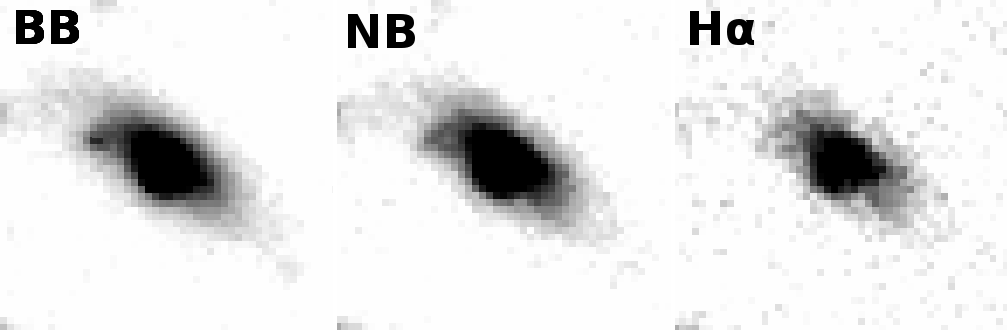}}
\fbox{\includegraphics[trim=0cm 0cm 0cm 0cm, width=0.46\textwidth]{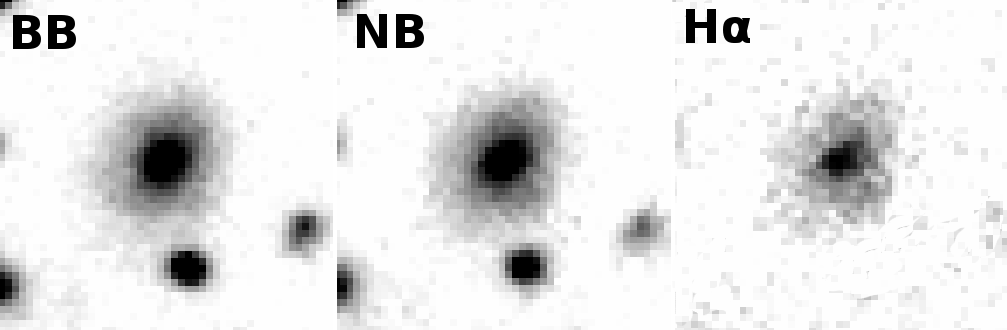}}
\end{center}
\caption{Eight examples of bright, extended H$\alpha$ emitters found in the proximity of the radio relics in the `sausage' cluster. Some examples show very disturbed, multi-tailed, asymmetric H$\alpha$ morphologies. Very few such examples were found in the `toothbrush' cluster. Within each group, the image to the left is the BB image, the central one is the NB and the right image is the BB subtracted from the NB. All images are on the same spatial scale. The red bar indicates a scale of $20$ arcsec, equivalent to $64$~kpc at the redshift of the `sausage' cluster. The circle in the emission-line image for the top-left emitter masks out an artifact created around an imperfectly NB-BB subtracted bright star. The images are for illustration purposes only as the subtraction, colour palette and contrast were chosen specifically to emphasise the distribution of ionised gas in the galaxies.}
\label{fig:emitters}
\end{figure*}

\subsection{Galactic dust extinction correction}
\label{sec:obs-reduction:dust}

As mentioned in \S\ref{sec:obs-reduction:obs}, both targets are affected by Galactic dust extinction. To correct for this effect, we linearly interpolated between SDSS i and z redenning values \citep{2011ApJ...737..103S} and used the fit to predict the Galactic extinction at the wavelength of our BB and NB filters. As the redenning varies significantly across the field, we produced `dust screens' across the FOV, instead of using single values (see Fig.~\ref{fig:dust} for NB extinction maps). The spatial resolution of the dust map is given by the IRAS survey map which was used to derive the dust extinction corrections ($4'$). The error on the dust extinction correction is of the order of $0.05$ mag. For this study, we are binning the sources based on their luminosity (i.e. magnitude) in bins which are much wider than the individual magnitude errors. Therefore random errors, such as the dust extinction correction error, are averaged out within a bin.

It should be noted that the varying dust extinction effect is also relevant for X-ray observations aiming at detecting shock fronts in galaxy clusters. Accurate subtraction of the background Galactic contribution to the number of photons is crucial for constraining shock parameters such as the Mach number. All X-ray observations of radio relics have taken a single off-cluster pointing which can lead to large biases in the derived model parameters \citep[e.g.][]{2013MNRAS.429.2617O, 2013MNRAS.433..812O, 2013PASJ...65...16A}.

\subsection{Narrow band excess selection}
\label{sec:obs-reduction:NBselection}
We follow the method of \citet{1995MNRAS.273..513B} \citep[see also][]{2009MNRAS.398...75S, 2012MNRAS.420.1926S} to select emission-line galaxies. In summary, we single-out emitters based on the colour excess significance of the narrow line with respect to the broad band emission ($\Sigma$) and the equivalent width (EW) of the line. We use these two cuts to model the scatter at the faint magnitudes and to reject bright sources with steep continuum that do not have an emission line. In our case, the colour significance $\Sigma$ is defined by:
\begin{equation}
\label{eq:Sigma}
\Sigma= \frac{10^{-0.4\left(m_{BB}-m_{NB}\right)}}{10^{-0.4(ZP_{AB}-m_{NB})} \sqrt{\pi r^2 \left(\sigma^2_\mathrm{NB}+\sigma^2_\mathrm{BB}\right)}},
\end{equation}
where $ZP_{AB}$ is the zero-point in magnitudes, $m_{NB}$ and $m_{BB}$ are the NB and BB magnitudes, respectively, $r$ is the radius of the aperture in pixels, $\sigma_\mathrm{NB}$ and $\sigma_\mathrm{BB}$ are the rms noise levels in the NB and BB images, respectively. The $\Sigma$ parameter is a signal-to-noise type of measurement that calculates the significance of the excess based on the RMS scatter of the intrinsic magnitudes at the faint end.

The EW is directly related to the $\mathrm{BB}-\mathrm{NB}$ colour through the emission line flux. The NB or BB flux $f_{NB,BB}$ depend on the magnitude by:
 \begin{equation}
\label{eq:flux_broad}
f_{NB,BB} = \frac{c}{\lambda^2_{NB,BB}} 10^{-0.4(m_{NB,BB}-ZP_{AB})},
\end{equation}
where $c$ is the speed of light and $\lambda_\mathrm{NB}$ and $\lambda_\mathrm{BB}$ are the central effective wavelengths of the two filters. The line flux can then be derived as:
\begin{equation}
\label{eq:flux}
F_{line} = \Delta\lambda_\mathrm{NB} \frac{f_{NB}-f_{BB}}{1-\Delta\lambda_{NB}/\Delta\lambda_{BB}},
\end{equation}
where $\Delta\lambda_\mathrm{NB}$ and $\Delta\lambda_\mathrm{BB}$ are the widths of the NB and BB filters. The EW is then:
\begin{align}
\label{eq:EW}
\mathrm{EW} & = \Delta\lambda_\mathrm{NB} \frac{f_{NB}-f_{BB}}{f_{BB}-f_{NB}\left(\Delta\lambda_{NB}/\Delta\lambda_{BB}\right)} \\
& = - \Delta\lambda_\mathrm{BB} \frac{1 - \left(\lambda^2_\mathrm{NB}/\lambda^2_\mathrm{BB}\right) 10^{-0.4\left(m_{BB}-m_{NB}\right)}}{1- \left(\Delta\lambda_\mathrm{BB}/\Delta\lambda_\mathrm{NB}\right)\left(\lambda^2_\mathrm{NB}/\lambda^2_\mathrm{BB}\right) 10^{-0.4\left(m_{BB}-m_{NB}\right)}}.
\end{align}
The observed EW relates to the intrinsic $\mathrm{EW}_0$ at emission via the redshift $z$:
\begin{equation}
\label{eq:EW0}
\mathrm{EW}_0 = \mathrm{EW}/\left(1+z\right).
\end{equation}

\begin{figure*}
\begin{center}
\includegraphics[trim=0cm 0cm 0cm 0cm, width=0.495\textwidth]{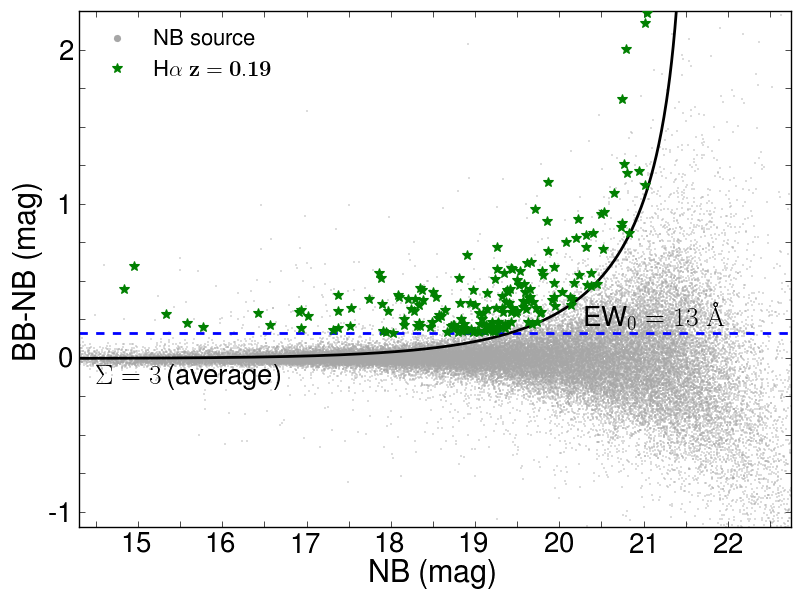}
\includegraphics[trim=0cm 0cm 0cm 0cm, width=0.495\textwidth]{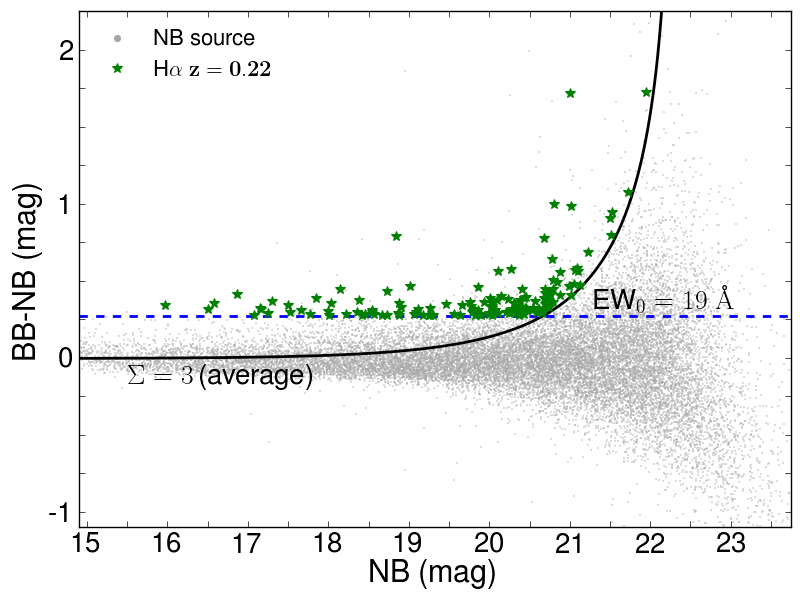}
\end{center}
\caption{Colour-magnitude diagrams displaying narrow-band excess as function of NB magnitude. \textit{Left}: `Sausage'. \textit{Right}: `Toothbrush'. Notice that due to variations in dust extinction across the FOV and dissimilar chip properties, the selection of emitters was performed separately for each chip, allowing for different rms values. The data were corrected for the colour depence on the NB magnitude. The curves represent the average $3\Sigma$ colour significances (for the average depth, as the analysis is done individually for each chip). The horizontal dashed lines are the rest-frame EW cuts used in this study.}
\label{fig:colmag}
\end{figure*}

\begin{table*}
\begin{center}
\caption{Sources detected in our survey. The detected number of sources are within $5$ arcsec diameter apertures down to $3\sigma$ magnitude limits within the NB matched to BB catalogues. The number of rejected stars, total number of NB emitters and extended NB emitters (sizes greater than $4$ arcsec) for the two fields are also given. The seventh column presents the number of emitters found in the proximity of the radio relics, based on the arc-sectors defined in Fig.~\ref{fig:image}. The penultimate column shows the number of extended emitters located nearby radio relics. The last column tabulates the total number of sources likely to the H$\alpha$ emitters at the redshift of the clusters. The total was calculated as the sum of the number of extended emitters plus a fraction of $\sim20$ of other emitters in the field (see \S\ref{sec:results:Halpha}).}
\begin{tabular}{l l c c c c c c c}
\hline
\hline
Field & Filter & Detected  & Flagged & Emitters  & Extended  & Emitters close  & Extended emitters & H$\alpha$-likely \\ 
& & ($3\sigma$) & stars & ($\Sigma>3$) & emitters & to the relics & close to the relics & sources \\ \hline
\multirow{2}{*}{`Sausage'} & NOVA782HA & 36196 & \multirow{2}{*}{357} & \multirow{2}{*}{181} & \multirow{2}{*}{33} & \multirow{2}{*}{31} & \multirow{2}{*}{13} & \multirow{2}{*}{49}  \\
 & WFCSloanI & 40007 &   & \\ \hline
\multirow{2}{*}{`Toothbrush'}  & NOVA804HA & 16776 & \multirow{2}{*}{207} & \multirow{2}{*}{141} & \multirow{2}{*}{12} & \multirow{2}{*}{6} & \multirow{2}{*}{1}  & \multirow{2}{*}{30} \\
 & WFCSloanI & 19418 &   & \\
\hline
\end{tabular}
\label{tab:sources}
\end{center}
\end{table*}

Because the NB filters do not fall at the centre of the BB filter there is a positive systematic offset of the colour excess ($\mathrm{BB}-\mathrm{NB}$). For the `sausage', a constant median offset provides a good description of the data. For the `toothbrush', we fit a linear regression to the non-saturated magnitudes at the bright end (NB magnitudes between $14.5$ and $18.5$) to correct the NB magnitudes and consequently the colour excess. The slope for the `toothbrush' cluster is caused by the non-central location of its associated NB filter inside the BB filter. That means more red galaxies have a stronger positive colour excess and may be selected as emitters. The NB magnitudes, and the excess colour consequently, were corrected by $-0.167$ for the `sausage'. NB magnitudes in the `toothbrush' were corrected by $0.023(\mathrm{BB}-\mathrm{NB})-0.510$ (see Fig.~\ref{fig:colmag}).

The EW cut is imposed to distinguish true emission-line systems from sources which have positive excess because of random scatter in the magnitude measurements, such as foreground stars, or steep continuum. Therefore, we measure the standard deviation of the colour distribution $\sigma_\mathrm{excess}$ around zero and consider only sources with $\mathrm{BB}-\mathrm{NB}$ larger than $3\sigma_\mathrm{excess}$. As shown in equation~\ref{eq:EW}, colour excess can be directly related to equivalent width \citep{2009MNRAS.398...75S}. The data obtained with the two filters have different scatterings of the excess values, i.e. the standard deviation of the `toothbrush' excess values is higher than for the `sausage'. Therefore, we choose separate EW cuts that reflect the statistical properties of the magnitude measurements of the two fields. For the `sausage' we use a colour cut at $0.16$ mag and for the `toothbrush' we use $0.27$. These correspond to an observed EW of $16$ {\AA} and an intrinsic $\mathrm{EW}_0$ at emission of $13$ {\AA} for the `sausage' and $\mathrm{EW}=24$, $\mathrm{EW}_0=19$ for the `toothbrush'. The higher scatter in the `toothbrush' is intrinsic to the filter and cannot be attributed to varying median colour offsets caused by different properties of the four chips. We also tested a single, common EW cut for the two clusters and found that a higher EW cut does not change the results.

A colour excess significance of $\Sigma>3$ was imposed such that we remove most spurious sources and obtain a robust sample of emitters. This ensures we reject sources which have a low signal-to-noise. Because of the low Galactic latitude of both our targets, the fields are extremely crowded with stars, including old stars and frequent double-star occurrences, uniformly distributed across the field of view. Because of the rather large apertures used for source detection, light from companions in double-star systems can contaminate the measurement. All of the NB H$\alpha$ imaging surveys to date have avoided observing in the Galactic plane to obtain samples as clean as possible. One example is the COSMOS field \citep{2007ApJS..172....1S, 2008ApJS..175..128S}, where the source density down to a magnitude of $22$ is $\sim5$ sources/arcmin$^2$. We detect $17$ sources/arcmin$^2$ in the `toothbrush' field, which is at Galactic latitude $9.0^{\circ}$, while for the `sausage', at latitude $-5.1^{\circ}$, we detect an average of $45$ sources/arcmin$^2$. The source density for the two fields discussed in the present work is significantly higher than for a typical H$\alpha$ NB target field. The difference can be accounted for by the dense star field in the line of sight towards the Galactic plane, so an inspection of the dataset is crucial in obtaining robust samples of emitters. A first pass removal of stars was performed based on the source ellipticities reported by {\sc sextractor}. We further performed two independent visual inspections of the images to flag potential false positives around the edges of the chip where the noise increases or close to bright, saturated stars, mismatched NB to BB emitters and double stars blended into a single source which mimic line emission. We cross-checked the sources labelled as stars by {\sc sextractor} and the visual inspection and very good agreement was found. 

Therefore, for a source to enter the emitters catalogue, it needs to fulfil three conditions (see~Fig.~\ref{fig:colmag}):
\begin{itemize}
\item colour excess higher than $3\sigma_\mathrm{excess}$ scatter at bright magnitudes (source has significant line emission)
\item $\Sigma>3$ colour excess significance (source has high signal-to-noise)
\item pass visual inspection
\end{itemize}
The sample of sources that fulfil all criteria consists of: $181$ emitters for the `sausage' and $141$ emitters for the `toothbrush'. We visually inspected all of the sources in the fields by looking at a difference image obtained by subtracting the BB image from the NB image. Our visual inspection of the images revealed that all obvious `sausage' line emitters pass our criteria. In the case of the `toothbrush', because of the higher EW cut, we miss a few extended emitters with low surface brightness, but high integrated fluxes. We note that we fully correct for the emitters that do not fulfil the criteria by doing a completeness study (see \S\ref{sec:results:completeness}). 

Upon visual inspection, we also noted that there were sources significantly larger than the bulk of the emitters and thus flagged sources larger than $4$ arcesc as extended. Examples of such bright, extended emitters (sizes greater than $4$ arcsec, equivalent to $\sim14$ kpc) are given in Fig.~\ref{fig:emitters} and they are predominantly found around the radio relics. We have to note that perfect subtraction of sources in the image plane is difficult because of non-matching point-spread-functions and different noise properties between the broad band and narrow band images. The $\mathrm{BB}-\mathrm{NB}$ images in Fig.~\ref{fig:emitters} are displayed with contrast and colour scheme chosen to enhance the H$\alpha$ gas distribution in the galaxies. We have displayed the distribution of extended line emitters in Fig.~\ref{fig:image}.

\section{RESULTS}
\label{sec:results}

\subsection{Selecting H$\alpha$ emitters}\label{sec:results:Halpha}
H$\alpha$ NB surveys at moderately low redshift can also detect galaxies with shorter rest-frame wavelength lines located at much higher redshifts. In our survey, we are also sensitive to H$\beta$ ($\lambda_\mathrm{rest}=4861$\;{\AA}) and [O{\sc iii}]$\lambda\lambda4959,5007$ emitters at $z\sim0.61-0.65$, and [O{\sc ii}] ($\lambda_\mathrm{rest}=3727$\;{\AA}) emitters at $z\sim1.15$. We also expect to detect 4000\;{\AA} break galaxies located over a wider range around $z=0.8$.

\begin{figure*}
\begin{center}
\includegraphics[trim=0cm 0cm 0cm 0cm, width=0.495\textwidth]{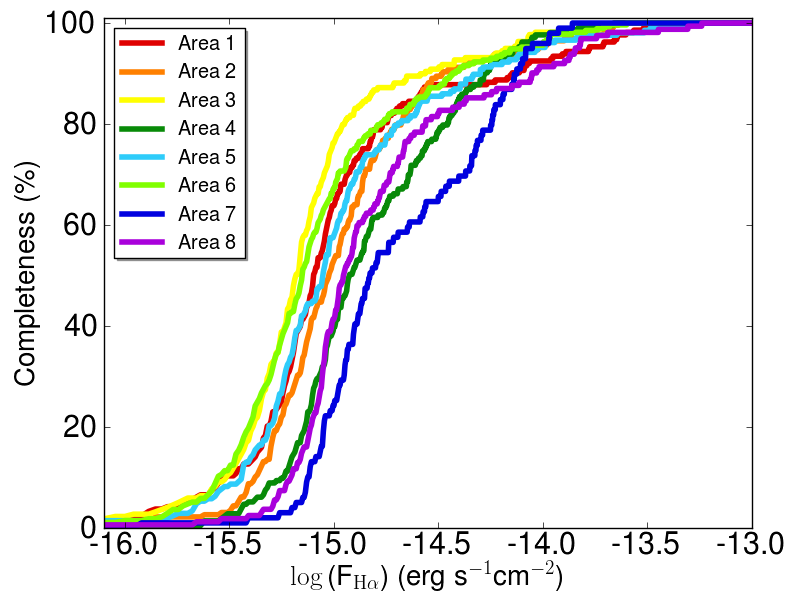}
\includegraphics[trim=0cm 0cm 0cm 0cm, width=0.495\textwidth]{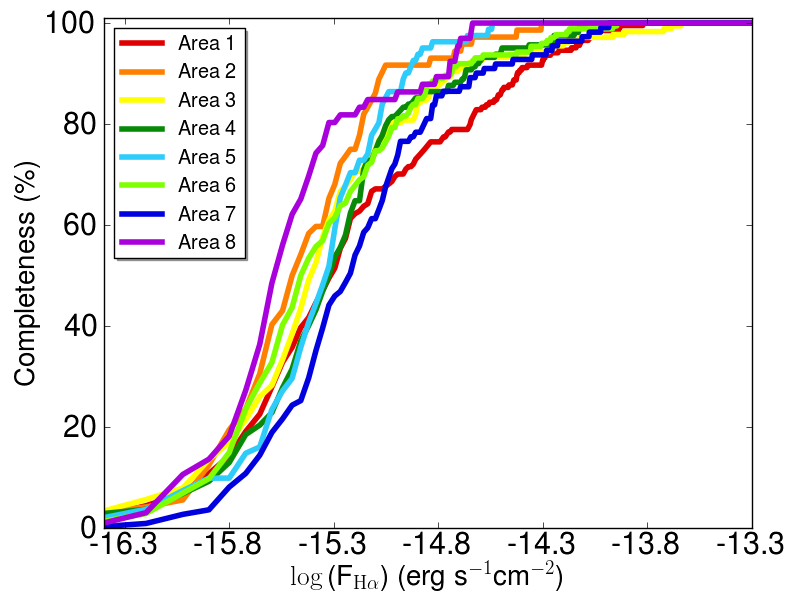}
\end{center}
\caption{Completeness of the surveys for the `sausage' and the `toothbrush' clusters as function of H$\alpha$ flux. The completeness study is performed separately on sub-areas of the FOV, as defined in Fig.~\ref{fig:dust}. The differences in the curves between the sub-areas arises from their different dust extinction properties. \textit{Left}: `Sausage'. \textit{Right}: `Toothbrush'.}
\label{fig:completeness}
\end{figure*}

Traditionally, when available, colour-colour information is used to separate higher redshift interlopers that fall within a NB filter designed to capture the H$\alpha$ emission \citep[e.g.][]{2008ApJS..175..128S,2013MNRAS.428.1128S}. While we do not have multi-band imaging of the two fields, we can derive and apply a statistical correction by making use of the COSMOS \citep{2007ApJS..172...99C, 2009ApJ...690.1236I} catalogue. \citet{2008ApJS..175..128S} derived catalogues of NB816 line-emitters and \citet{2013MNRAS.428.1128S} observed the same field with the narrow band filter NB921. We predicted broad-band $5$ arcsec magnitudes using their CFHT i magnitudes and measured narrow-band magnitudes in the COSMOS NB816 filter at $8160$\;{\AA} and NB921 at $\lambda=9120$\;{\AA}. In order to simulate the conditions of our two cluster catalogues, we made use of their ancillary z' filter information to apply a colour correction to the data. This is necessary because the NB816 filter is non-centrally located within the BB filter. We fully corrected for the colour dependence to best simulate the `sausage' field where the NB filter is placed very close to the centre of the BB filter. We applied a partial colour correction to the COSMOS magnitudes to obtain a catalogue slightly affected by colour dependence as in the case of the `toothbrush' dataset. We then passed the `sausage'-like and the `toothbrush'-like COSMOS catalogues through the same selection criteria presented in \S\ref{sec:obs-reduction:NBselection}. In this way, we fully mimic our data selection process. We then combined the NB816 emitters with the NB912 emitters from the \citet{2013MNRAS.428.1128S} sample to minimise cosmic variance. The photometric redshift information from the COSMOS catalogue was used in conjunction to the selection criteria to classify the potential emitters as H$\alpha$, H$\beta$+[O{\sc iii}] emitters, 4000\;{\AA} break galaxies or [O{\sc ii}] emitters. We used this information to derive H$\alpha$ fractions relative to the total number of emitters as function of flux (see Fig.~\ref{fig:fractions}) and correct the H$\alpha$ luminosity function. Typical H$\alpha$ fractions are around $20$ per cent. The expressions of the fractions $\mathrm{frac}_{H\alpha}$ as function of flux are given below. We have explored functions of different order, but these provided the best description of the data. For the `sausage' cluster:
\begin{align}
\label{eq:frac1}
\mathrm{frac}_{H\alpha} & = 2.264 \log^4 \left(F_{line}\right) + 144.2 \log^3\left(F_{line}\right) \notag \\
						& + 3441 \log^2 \left(F_{line}\right) + 3.646 \times 10^4 \log\left(F_{line}\right) \notag \\
						& + 1.447 \times 10^5,
\end{align}
where $F_{line}$ is the line flux. For the `toothbrush' field, the functional form is:
\begin{align}
\label{eq:frac2}
\mathrm{frac}_{H\alpha} & = 3.945 \log^4\left(F_{line}\right) + 248.1 \log^3 \left(F_{line}\right) \notag \\ 
& + 5843 \log^2 \left(F_{line}\right) + 6.105 \times 10^4 \log\left(F_{line}\right) \notag \\
& + 2.389 \times 10^5.
\end{align}
Extended sources (larger than $4$ arcsec) are very likely to be real H$\alpha$ emitters at the redshift of the cluster (see also \S\ref{sec:discussion}, where we detail why this is so). However, for sources which are smaller, size cannot be used as an argument to derive the redshift. Therefore, for these smaller sources, we cannot say, on a case by case basis, which are H$\alpha$ at redshift $\sim0.2$ and which are H$\beta$+[O{\sc iii}] emitters, 4000\;{\AA} break galaxies or [O{\sc ii}] emitters. We therefore apply the fractions derived above, which, in a statistical sense provide the correct number of counts for each luminosity bin, within the errors. We would like to stress that even though \citet{2008ApJS..175..128S} have separated the H$\alpha$ from all others emitters and then computed the LF, the final statistical result is the same as in our method.

Fig.~\ref{fig:fractions} reveals the increase of the H$\alpha$ fraction with luminosity. Therefore, applying a constant fraction across all luminosity ranges is unsuitable. Studies performed with the Subaru telescope and long integration times, such as those by \citet{2008ApJS..175..128S}, \citet{2007ApJ...657..738L} and \citet{2013MNRAS.433..796D}, capture the fluxes where H$\alpha$ contributes to less than $15$ per cent and they become saturated when the H$\alpha$ fraction is rising.

\subsection{Removing [N{\sc ii}] contamination}
The NB filters are wide enough to also capture the adjacent [N{\sc ii}] emission. It is crucial that we attempt to remove the contribution of this emission line from our fluxes. If unaccounted for, this contamination artificially increases the line fluxes and EWs. We follow the relation empirically calibrated by \citet{2012MNRAS.420.1926S} against a large SDSS sample to remove the [N{\sc ii}] contribution. The relationship takes a functional form between the logarithm of the combined EW of the H$\alpha$ and [N{\sc ii}] lines and the logarithm of the fractional contribution of [N{\sc ii}] towards the blended line flux:
\begin{equation}
\label{eq:NII}
f=-0.924+4.802E-8.892E^2+6.701E^3-2.27E^4+0.279E^5,
\end{equation}
where $f$ is the logarithmic ratio of [N{\sc ii}] flux to the total flux and $E=\log_{10}(\mathrm{EW}_0(\mathrm{H}\alpha+[\mathrm{N}\textsc{ii}]))$. The average contamination by [N{\sc ii}] flux for our sample is $0.31$ of the total blended flux.

\begin{figure*}
\begin{center}
\includegraphics[height=0.970\columnwidth]{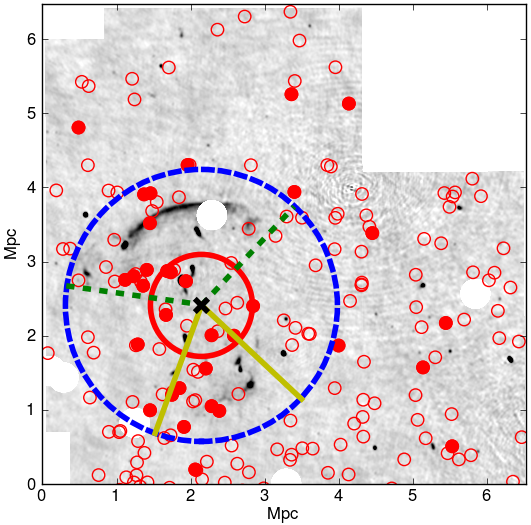}
\includegraphics[height=0.970\columnwidth]{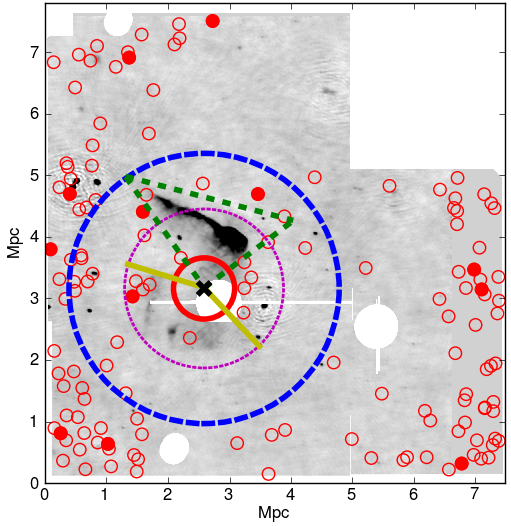}
\end{center}
\caption{Giant Meterwave Radio Telescope 323~MHz radio images in grey intensity. \textit{Left}: `Sausage' FOV. \textit{Right}: `Toothbrush' FOV. The figure also shows the FOV of the two fields with masked areas in white. Circular masked areas are masked bright stars. The red circles show the position of the line emitters detected in the narrow band study. The filled circles indicate the location of the extended emitters (sizes greater than 4 arcsec), likely to be H$\alpha$ emitters. Note the some source markers are overlapping. The arc-sectors define the areas around the radio relics which were considered for producing relic luminosity functions. The cluster centres are defined to be at the location of the black crosses. \textit{Left}: The Northern relic was captured by a section between the solid red and the dashed blue circles, bound by the dashed green lines. The Southern relic area was defined between the solid red and the dashed blue circles and the solid yellow radii. \textit{Right}: The Northern relic was captured by a section between the solid red and the dashed blue circles, bound by the dashed green lines. The Southern relic area was defined between the solid red and dotted purple circles and the solid yellow radii.}
\label{fig:image}
\end{figure*}

\subsection{H$\alpha$ luminosity}
Using emission line flux of a source, we compute its H$\alpha$ luminosity. The luminosity is calculated as:
\begin{equation}
\label{eq:L}
L_{\mathrm{H}\alpha}=4 \pi d^2_{L} F(H\alpha),
\end{equation}
where $F(H\alpha)$ is the flux and $d_{L}$ is the luminosity distance calculated assuming all sources are at redshift $0.1945$ or $0.2250$, for the `sausage' or the `toothbrush' respectively (see Table~\ref{tab:filters}). 

\subsection{Completeness correction}\label{sec:results:completeness}
For the purpose of building luminosity functions, we bin the sources based on their luminosity. It is essential to correct for incompleteness at the faint end of the source counts. Not doing so will result in a synthetically flat faint-end slope or even a turn-over in the LF. The completeness of the survey is determined from a set of simulations using the measured data itself. We follow the method developed by \citet{2012MNRAS.420.1926S, 2013MNRAS.428.1128S}. First, the distribution of recovered emitters as function of NB magnitude is modelled. A sub-sample is drawn from the pool of sources consistent with being non-emitters, following the same NB distribution as the population of emitters. Pure H$\alpha$ emission line fluxes are added to the sources and the sample is passed through the selection criteria described in \S\ref{sec:obs-reduction:NBselection}. Because of the irregular dust extinction (especially for the `sausage') and the slightly different properties of the four WFC chips, we divided the FOV into eight areas (see~Fig.\ref{fig:dust}), for which we independently study the completeness properties. The `extragalactic' depth of the observations is superior for areas with low dust extinction, while high-dust extinction areas will have a higher survey limit. This poses issues for computing the LF. Dividing the number counts detected over a restricted area with lower dust extinction by the entire FOV volume can simulate a turn-over in the faint-end slope of the LF. The results of the completeness study are presented in Fig.~\ref{fig:completeness}. As expected, the curves for the eight sub-areas are very similar for the `toothbrush' where the dust extinction is almost constant across the field. The highly variable `dust screen' for the `sausage' is reflected in the completeness curves for the different sub-areas: the completeness for two cells with different dust properties can differ by as much as $50$ per cent. We therefore correct the LF adaptively for the eight sub-areas. Sources whose flux falls below the $30$ per cent completeness limit are disregarded and their associated comoving volume is not taken into account.

\subsection{Volume}
Both our NB filters have a width of $110$ {\AA}. Using a perfect top-hat (TH) approximation for the NB filters, at FWHM this translates to an H$\alpha$ redshift coverage from $0.1865$ to $0.2025$ for the `sausage' and to $0.2168-0.2328$ for the `toothbrush'. The luminosity distance at the peak of the transmission curve is $936.2$ Mpc and $1116.1$ Mpc for the `sausage' and the `toothbrush', respectively. We survey a comoving volume density of $1.227\times10^4$~Mpc$^3$deg$^{-2}$ and $1.600\times10^4$~Mpc$^3$deg$^{-2}$, respectively. Taking into account the masked edges and saturated stars (see Fig.~\ref{fig:image}), we are surveying an effective comoving volume of $3.371\times10^3$~Mpc$^3$ for the `sausage' and $4.546\times10^3$~Mpc$^3$ for the `toothbrush'. Our volumes are comparable ($\sim 1/3$) to the volumes of field surveys such as those by \citet{2007ApJ...657..738L} ($\sim1.4\times10^4$~Mpc$^3$) and \citet{2013MNRAS.433..796D} ($\sim1.2\times10^4$~Mpc$^3$).

\subsection{Filter profile correction}
The volume covered by our NB filters is $90$ per cent of that covered by an idealised TH profile with the same FWHM and maximum equal to the peak of our NB filters. Because neither NB filter is a perfect TH (Fig.~\ref{fig:transmittance}), bright emitters will be detected in the wings of the filter profile as fainter sources. In order to estimate the magnitude of this bias effect, we performed a series of simulations based on the method of \citet{2009MNRAS.398...75S, 2012MNRAS.420.1926S}. This entails selecting emitters using a perfect TH filter and computing a first pass LF. Simulated H$\alpha$ emitters are generated using the resulting best-fit Schechter function. This population is then folded through the true filter profile and the resulting LF is compared with the idealised one to study the rate of recovery of emitters.

\begin{figure*}
\begin{center}
\includegraphics[width=0.47\textwidth]{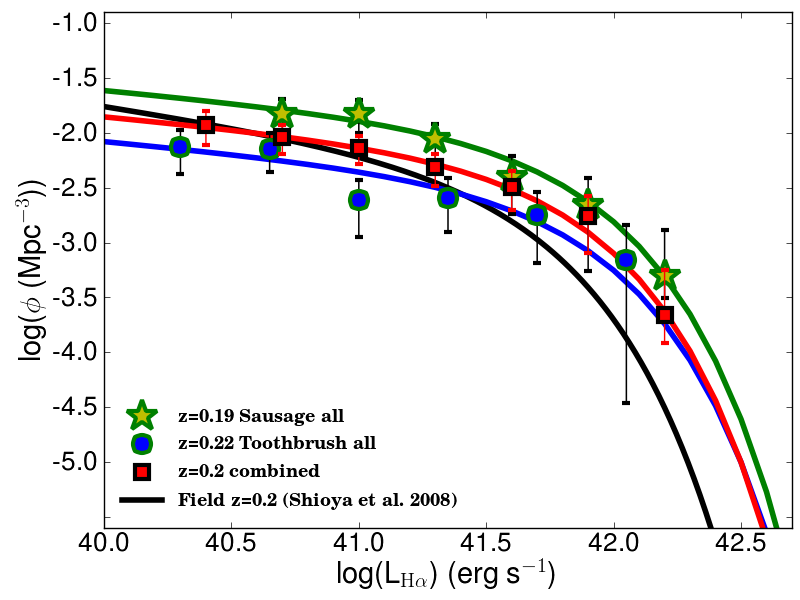}
\includegraphics[width=0.47\textwidth]{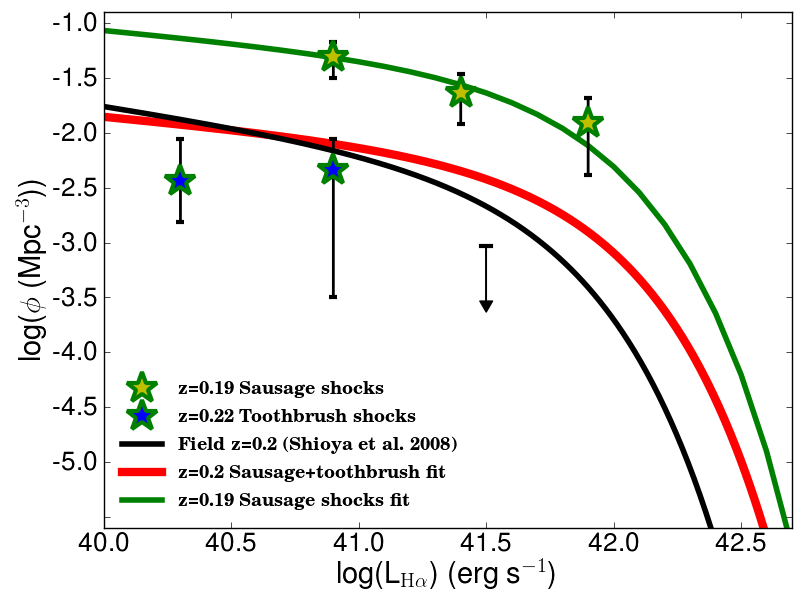}
\end{center}
\caption{Luminosity function for the two clusters. \textit{Left}: Luminosity functions using emitters from the entire FOV. The best-fitting parameters of the LFs are given in Table~\ref{tab:LF}. We combined the data for the two clusters to obtain an average field-of-view wide LF. Overplotted is the LF derived by \citet{2008ApJS..175..128S} for redshift $0.24$. \textit{Right}: The H$\alpha$ LF for the regions projected around the `sausage' and the `toothbrush' cluster relics. The regions used are defined in Fig.~\ref{fig:image}. Overplotted is the average field-of-view LF derived in the present work. An upper limit of detection is shown for the `toothbrush' cluster, corresponding to $1$ source per bin. A fit was produced to describe the `sausage' emitters nearby the two relics by fixing the $L^*$ and $\alpha$ parameters to the value of the average LF and varying only $\phi^*$, as this produces a very good fit to the data.}
\label{fig:LF}
\end{figure*}

\subsection{Survey limits}\label{sec:results:limits}
Down to a $30\%$ completeness limit, our LFs probe down to $10^{40.6}$~erg~s$^{-1}$ for the `sausage' and $10^{40.4}$~erg~s$^{-1}$ for the `toothbrush'. By using the relationship derived by \cite{1998ARA&A..36..189K} corrected for the initial mass function from \cite{2003PASP..115..763C}, we can obtain a limiting star formation rate of our survey of $0.17$ and $0.11$ M$_{\sun}$ yr$^{-1}$ for the `sausage' and the `toothbrush', respectively. Our datasets become saturated at magnitude $14.5$, which means we can probe up to luminosities of $10^{43.2}$~erg~s$^{-1}$. 

We should note that comparison to NB H$\alpha$ surveys of blank fields such as COSMOS or the UKIDSS Ultra Deep Survey Field can be challenging. These studies have been carried out with wide-aperture telescopes using long individual integration times designed to target high-redshift Ly$\alpha$. This means the data becomes saturated at magnitudes brighter $10^{41.0\sim41.5}$~erg~s$^{-1}$ and that bright H$\alpha$ emitters can be missed \citep[e.g.][]{2008ApJS..175..128S, 2013MNRAS.433..796D}. These results are biased against high luminosities, which leads their fits to underpredict the high-luminosity behaviour of the LF. Therefore, luminosity functions derived from lower luminosity data cannot be easily translated to our luminosity coverage. This strengthens the importance of using smaller-diameter telescope for NB studies and/or short individual exposures that are not severely saturated at moderate magnitudes.

\subsection{H$\alpha$ luminosity function}
We bin the data based on their luminosity to build LFs. We use the entire sample of H$\alpha$ emitters, after applying all the relevant corrections (statistical removal of other line emitters, volume, completeness and filter profile corrections), to obtain a robust luminosity function of the field-of-view of the two clusters. We fit the data with a Schechter function \citep{1976ApJ...203..297S} of the form:
\begin{equation}
\label{eq:schechter}
\phi(L) \mathrm{d} L = \phi^*\left(\frac{L}{L^*}\right)^{\alpha} e^{-(L/L^*)} \mathrm{d} \left(\frac{L}{L^*}\right).
\end{equation}

\begin{table*}
\begin{center}
\caption{Schechter fit parameters according to equation~\ref{eq:schechter} for the `sausage' and the `toothbrush' clusters at redshift $\sim0.2$. The data for the two fields was combined and fitted separately  with a Schechter function for comparison. All errors represent $1\sigma$ uncertainties estimates on the fitted parameters. The uncertainty in $\alpha$ was obtained by keeping all parameters free and minimising the $\chi^2$ of the fit. We then fixed $\alpha$, and refitted the Schechter functions with $\log \phi^*$ and $\log L^*$ as free parameters. The errors on $\log \phi^*$ and $\log L^*$ are $1\sigma$ uncertainties determined from Schechter with two free parameters. For the fit on the H$\alpha$ emitters located around the relic area within the `sausage' cluster, we fixed both $\alpha$ and $\log L^*$ and varied the normalisation. The fit from \citet{2008ApJS..175..128S} derived for a blank field at a similar redshift is given as reference (after being uncorrected for `extragalactic' H$\alpha$ dust extinction). The last column gives the number of emitters employed in building the LFs. The first value tabulates extended emitters likely to be at the redshift of the clusters, while the second value gives the number of other sources, which could be H$\alpha$ or other, higher-redshift line emitters. For this second group of sources, a fraction was applied to account for the actual percentage we expect to be H$\alpha$ at the redshift of the cluster (see eq.~\ref{eq:frac1} and~\ref{eq:frac2}).}
\begin{tabular}{l l c c c}
\hline
\hline
Field                       & $\alpha$ & $\log \phi^*$ & $\log L^*$ & Number of H$\alpha$ \\ 
							 &			& Mpc$^{-3}$ &  erg s$^{-1}$ & (extended+other)            \\\hline  \vspace{3pt}
`Sausage'                   & $-1.2$    &  $-2.33^{+0.11}_{-0.11}$   & $41.73^{+0.08}_{-0.06}$      & $33+16$  \\ \vspace{3pt}
`Toothbrush'                & $-1.2$    &  $-2.82^{+0.16}_{-0.31}$   & $41.77^{+0.91}_{-0.21}$  & $12+18$   \\ \vspace{3pt}
Combined					 & $-1.2^{+0.2}_{-0.3}$    &  $-2.57^{+0.09}_{-0.09}$   & $41.72^{+0.11}_{-0.07}$ & $45+50$ \\ \vspace{3pt}
`Sausage' relic area        & $-1.2$    &  $-1.77^{+0.09}_{-0.09}$   & $41.72$\phantom{00000} & $13+2$\phantom{0} \\ \vspace{3pt}
\citet{2008ApJS..175..128S} & $-1.35^{+0.11}_{-0.13}$    &  $-2.65^{+0.27}_{-0.38}$   & $41.57^{+0.38}_{-0.23}$ &  \\
\hline
\end{tabular}
\label{tab:LF}
\end{center}
\end{table*}

We combine the two datasets to obtain an average field-wide LF down to a limiting luminosity of $>10^{40.5}$~erg~s$^{-1}$ (limiting SFR of $0.14$). In the fit, we consider sources with line fluxes above the $30\%$ completeness limit. We note that extended sources, likely to be at the redshift of the clusters, were assigned a weight of $1$. For other line emitters, because of the ambiguity in the nature of the line/feature (H$\alpha$, H$\beta$+[O{\sc iii}], 4000\;{\AA} break galaxies or [O{\sc ii}]), the H$\alpha$ fractions derived in \S\ref{sec:results:Halpha} are applied (to recover the correct statistical number of H$\alpha$ emitters). We first kept all three parameters free and, via a $\chi^2$ minimisation scheme, found that the best description of the data is given by a faint-end slope of $\alpha=-1.2^{+0.2}_{-0.3}$. This value is placed between the results of \citet{2013MNRAS.433..796D} ($-1.03^{+0.17}_{-0.15}$) and \citet{2008ApJS..175..128S} ($-1.35^{+0.11}_{-0.13}$) at similar redshift. \citet{2007ApJ...657..738L} obtain an even steeper value of $\alpha=-1.7 \pm 0.1$. Values obtained by \citet{2008ApJS..175..128S} for the COSMOS blank field at a similar redshift of $\sim0.24$ are given for reference \citep[see also:][]{2010ApJ...712L.189D, 2013MNRAS.433..796D}. One can immediately notice that the average field-wide LF matches the field data points obtained by \citet{2008ApJS..175..128S}. This is expected, since by combining two distinct datasets we minimise the effects of cosmic variance, as well as the contribution of the cluster environment to the overall shape of the luminosity function. Note the differences at the bright end: these are likely driven by Shioya's data being saturated at these luminosities (see \S\ref{sec:results:limits}). We do not observe changes in the faint-end slope $\alpha$, although there are hints of flattening compared to the canonical $-1.35$. To confirm this point, we also inspect LFs derived from the field outside the cluster areas (top and right CCDs, resulting in $69$ emitters or $18$ H$\alpha$-likely for the `sausage' and $77$ or $15$ H$\alpha$ for the `toothbrush') and conclude that the LFs move in the direction of matching the \citet{2008ApJS..175..128S} field LF, as is expected from cosmic variance.

We then fix $\alpha$ to $-1.2$ and vary only the normalisation and $L^*$. We fit the data for two cluster field-of-views with Schechter functions down to an observed H$\alpha$ luminosity of $>10^{40.6}$~erg~s$^{-1}$ for the `sausage' and $>10^{40.4}$~erg~s$^{-1}$ for the `toothbrush'. For this step, we use all emitters found across the entire field mapped by the WFC as shown in Fig.~\ref{fig:image}. We obtain LFs for each cluster field-of-view by using $\chi^2$ minimisation with two free parameters ($\phi^*$ and $L^*$). The results with $1\sigma$ uncertainties are summarised in Table~\ref{tab:LF} and seen in Figure~\ref{fig:LF} (left panel). Errors of data points are Poissonian. The overall normalisation of the `sausage' field LF ($\phi^*$) is significantly higher than the field LF, while the `toothbrush' seems to be underdense as compared to \citet{2008ApJS..175..128S} fit. Nevertheless, such differences are expected due to cosmic variance. 

\subsubsection{Shock areas}
In order to study the effect of the travelling shock waves in shaping the H$\alpha$ luminosity function of the cluster galaxies we define circular areas that trace the shock fronts in a shell-like pattern. The areas can be visualised in Fig.~\ref{fig:image}. As mentioned in \S\ref{sec:intro}, both clusters probably result from a major merger between two massive clusters, which in turn produces outward-travelling spherical shock waves. This motivates the choice of symmetrical, sector-of-arc regions to capture the relic areas. Both clusters host a strong Mach $M\sim4.5$, $\sim1.5$~Mpc-wide shock front at their northern outskirts traced by the dominating radio relic. Towards the south and east of the `sausage', there are smaller, patchier relics which probably trace broken shock fronts \citep{2013A&A...555A.110S}, where turbulence leads to mixing of different electron populations. Our samples of emitters in the proximity of the relic areas contains $31$ sources for the `sausage' cluster ($15$ likely to be H$\alpha$) and $6$ for the `toothbrush' cluster ($2$ likely to be H$\alpha$). We note that these samples are derived over comparable volumes and we thus find that the `sausage' has a density of H$\alpha$ emitters nearby relics $\approx7.5$ times higher than the equivalent area for the `toothbrush'.

The measurements of the `toothbrush' H$\alpha$ emitters in the proximity of shock fronts are consistent with field properties. The non-detection at the high-luminosity end is significant: an upper limit corresponding to $1$ source detected in the surveyed volume indicates that there are no emitters brighter than $10^{41.5}$~erg~s$^{-1}$. However, the `sausage' LF is in stringent tension with the field properties, being boosted by an order of magnitude with respect to a blank field, as well as our combined field-wide LF. We fit the `sausage' relic data with a Schechter function with $\alpha$ and $\log L^*$ fixed to the values we obtained from our combined fit (see Table~\ref{tab:LF}). This allows us to obtain a value of the normalisation of the LF $0.8$dex higher than what is found for the field (equivalent to a significance level of 9$\sigma$).

\subsubsection{Comparison to other clusters}
There are few NB studies of the H$\alpha$ luminosity function of clusters at $z\approx0-0.2$. We compare our LF for the cluster areas projected onto the vicinity of the relics with the results from cluster Abell 521 at redshift 0.25 \citep{2004ApJ...601..805U} and Abell 1367 and Coma at $z=0.02$ \citep{2002A&A...384..383I}. For this purpose, we follow the method of \citet{2004ApJ...601..805U} and \citet{2002A&A...384..383I} and estimate the cluster volume as a sphere with radius equal to projected size of the cluster on the sky (2 Mpc). In order to be fully consistent with the cluster studies of \citep{2002A&A...384..383I} and \citep{2004ApJ...601..805U}, we assume all emitters within the relic areas to belong to the cluster. We renormalise the line-emitter number counts to this estimated cluster volume, instead of using the entire surveyed volume. To produce comparable results, we also remove the $A_{H\alpha}$ correction for `extragalactic' H$\alpha$ extinction from the LFs of \citet{2004ApJ...601..805U} and \citet{2002A&A...384..383I}. 

\begin{figure}
\begin{center}
\includegraphics[width=0.47\textwidth]{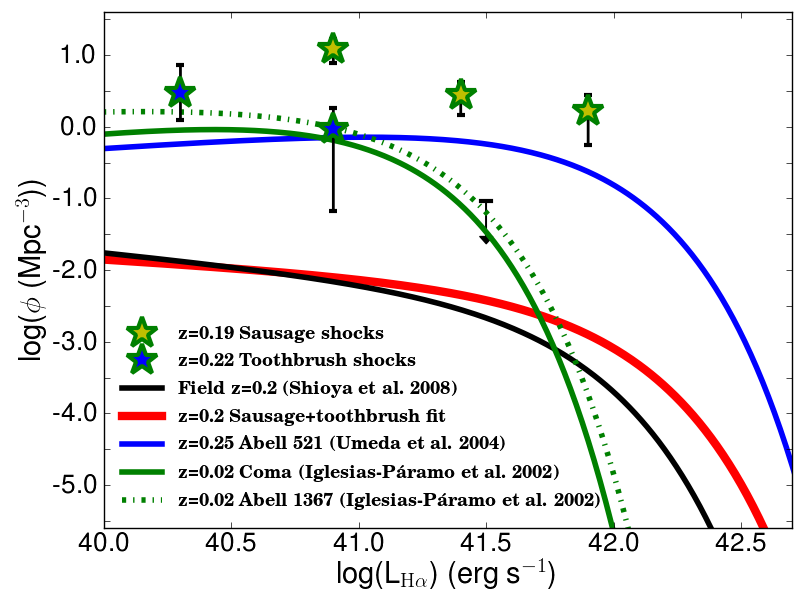}
\end{center}
\caption{H$\alpha$ narrow-band luminosity functions for clusters. The H$\alpha$ luminosity function for the regions projected around the relics in the `sausage' and the `toothbrush' clusters are shown as stars. An upper limit of detection is shown for the `toothbrush' cluster, corresponding to $1$ source within the surveyed volume. Note that here we have calculated the LF of the `sausage' and `toothbrush' clusters using a cluster volume corresponding to a sphere of radius of $2$ Mpc in order to compare to other cluster studies. Overplotted are the average cluster LF derived in the present work by using the entire cluster and field volume. We also plot the $z=0.2$ field LF from the work of \citet{2008ApJS..175..128S}. We also plotted LFs for cluster Abell 521 at $z=0.24$ \citep{2004ApJ...601..805U} and local clusters Abell 1367 and Coma at $z=0.02$ \citep{2002A&A...384..383I}.}
\label{fig:LFcluster}
\end{figure}

We plot the resulting bins for the `sausage' and the `toothbrush' clusters on top of the LFs derived for Abell 521, 1367 and Coma in Fig.~\ref{fig:LFcluster}. The overall LF normalisation for all clusters is above the field measurement of \citet{2008ApJS..175..128S} by several orders of magnitude. The measurements for the `toothbrush' relic areas now fall on the LF for local clusters at $z=0.02$, while our `sausage' measurements are in stringent disagreement with the local cluster results and an order of magnitude higher than the Abell 521 LF (at the 9$\sigma$ level).

\section{DISCUSSION}\label{sec:discussion}

We have carried out the first wide-area H$\alpha$ survey of two merging galaxy clusters that host radio relics. These Mpc-wide synchrotron structures trace twin-outward travelling shock waves that accelerate ICM electrons. Being interested in the possible effects the shock front has on the member galaxies, we study their properties, with a particular focus on the general cluster H$\alpha$ luminosity function and the H$\alpha$ LF around the relic areas. 

We used all of the potential H$\alpha$ emitters in the field of view to derive general LFs. We corrected for variable Galactic dust extinction, incompleteness, volume, filter profile and statistically corrected for the H$\alpha$ fraction. The `toothbrush' data is consistent with the field results derived by \citet{2008ApJS..175..128S} for the blank COSMOS field, while the `sausage' field LF indicates a larger number of bright emitters. 

The LF of H$\alpha$ galaxies around the relics in the `sausage' is significantly boosted compared to the field around the cluster, as well as the COSMOS field \citep{2008ApJS..175..128S} at a significance of $9\sigma$. The effect becomes more pronounced if we compare to the LFs by \citet{2013MNRAS.433..796D} or \citet{2007ApJ...657..738L}. Particularly important to note here is that our narrow band filters are wide enough to probe the background and foreground field behind and in front of the clusters, respectively. Our filters probe $30$ Mpc behind and $30$ Mpc in front of the cluster centre. The virial radius of the largest, richest clusters such as the Coma cluster is $2$ Mpc, while clusters like ours are expected to be comparable in size. Therefore, a small percentage of the volume we are probing is populated by the cluster (up to $10$ per cent), while the rest is foreground and background field. 

Excluding the volume not contained in the clusters themselves would lead to an increase in the normalisation of the LF by a factor of $10$ and would further strengthen our results. We attempt to account for this and follow the method of \citet{2002A&A...384..383I} and \citet{2004ApJ...601..805U} and approximate the `sausage' and `toothbrush' clusters to be contained within a spherical volume of radius 2~Mpc. This enables a comparison to cluster studies of Abell 521, Abell 1367 and Coma (see Fig.~\ref{fig:LFcluster}). The results for the relic areas in the `sausage' cluster are an order of magnitude higher than the LF derived for the three clusters Abell 521, 1367 and Coma. The normalisation $\phi^*$ of Abell 1367, Coma and Abell 521 is $\sim2$ orders of magnitude above the field LF at redshift $0.2$. For the relaxed, local clusters Abell 1367 and Coma the characteristic luminosity $L^*$ is below the field, which indicates a lack of bright emitters within the cluster volume, whereas the Abell 521 LF predicts the presence of many luminous emitters. \citet{2004ApJ...601..805U} suggest this enhancement can be attributed to the merging nature of the cluster. \citet{2008A&A...486..347G} have found that Abell 521 also hosts a radio relic, but without a distribution of H$\alpha$ emitters, we cannot confirm whether the majority of the emitters are located around the relic areas (\citet{2004ApJ...601..805U} do not present a distribution of their H$\alpha$ emitters). The measurements for the `sausage' cluster indicate a possible LF shape similar to that of Abell 521, driven by the merging nature of the cluster, but with a factor of $10$ increase in the overall normalisation. 

The visual inspection of the BB, NB and BB-NB subtracted images revealed the same effect (see Fig.~\ref{fig:image}). Around the smaller relics in the `sausage' cluster, where we expect the weaker, broken shock fronts to reside, we find numerous galaxies with haloes extended up to $60$~kpc H$\alpha$ (for some examples, see Fig.~\ref{fig:emitters}). Some of these emitters are highly disturbed and asymmetric and present tailed, clumpy H$\alpha$ morphologies. We do not find any such examples in the `toothbrush' cluster, nor in the field around the cluster. Sources extended beyond $4$ arcsec were classified as extended, which corresponds to $14$ kpc at the redshift of the clusters. Were these sources higher-redshift emitters, then their linear sizes would be extremely large: H$\beta$ and [O{\sc iii}] emitters at $z\sim0.61-0.65$ would measure $27$ kpc, 4000\;{\AA} break galaxies around $z=0.8$ would be $30$ kpc wide and [O{\sc ii}] emitters at $z\sim1.15$ would be $33$ kpc. A typical physical diameter of star-forming galaxies/line emitters at these redshift is $7-8$ kpc, with the largest galaxies being $\sim15$ kpc in diameter \citep{2013MNRAS.430.1158S}. This would render our extended emitters, were they at high redshift, two times larger than the largest galaxies and at least five times larger than typical galaxies at their redshift.

It has been known since the 1970s \citep{1972Natur.237..269M} that the cluster galaxies interact strongly with their environment: the ICM ram pressure strips the gas from the galaxies and bends their radio jets in a head-tail or twin-tail morphology. From our studies of the `sausage' cluster \citep{2013A&A...555A.110S}, we have discovered a clear bimodal distribution of galaxies with radio morphological orientations indicative of two sub-clusters, to the North and South of the cluster centre, moving in opposite directions, behind the radio shock fronts (for definition of cluster centres, see Fig.~\ref{fig:image}; the cluster centre was chosen as the middle of the distance between the radio relics, on the peak of the X-ray emission). The radio morphologies of the sources can be visualised in Fig.~\ref{fig:image}. Simulations suggest both clusters are post-core passage, meaning that the two initial clusters that participated in the merger, under the influence of gravity, have fallen into and passed through each-other once \citep{2011MNRAS.418..230V, 2012MNRAS.425L..76B}. Because both clusters are post-core passage mergers, the shock fronts have travelled more than $1$ Mpc though the ICM, potentially interacting with the cluster members. It is conceivable that travelling shock fronts could have an extra effect on the morphology of the galaxies, particularly on their H$\alpha$ haloes.  

We speculate that the disturbance of the `sausage' H$\alpha$ haloes and the increased number of bright H$\alpha$ emitters may be an effect of shock induced star-formation. Most of the galaxies in the cluster have interacted with the travelling shock wave sometime since core passage ($\sim1$~Gyr). The shock front passage effectively compressed and injected turbulence into the ionised gas. The results indicate that the shock front has not stripped the ionised materials from the host galaxy, but has compressed the gas, which then may have collapsed into star forming clouds. A similar scenario has been proposed for high redshift galaxies, albeit with a different source and scale for the shock front. In the case of high-redshift radio galaxies the alignment of the optical and radio continuum emission was interpreted by \citet{1989MNRAS.239P...1R} as shock induced star formation. These objects are still in the process of formation and their radio lobes would be advancing at Mach numbers of $M\simeq10-100$. Such high Mach numbers are difficult to attain in the context of cluster mergers, where they are limited by the relative speeds of the infalling clusters driven by gravitational attraction. Lower Mach numbers are pervasive in merging clusters and this is exemplified by the radio relics in the `sausage' cluster which trace shock fronts with Mach numbers between $\sim2-4.5$. Simulations of merging clusters show that the travelling shock fronts have higher Mach numbers in the past, compared to the shock strengths we measure via the relic radio properties \citep[e.g.][]{1993ApJ...407L..53R,2011MNRAS.418..230V}. It is therefore conceivable that large scale shocks with currently observed low Mach number could induce Jeans instabilities in the galactic gas clouds and cause them to collapse into star formation.

The boost in the H$\alpha$ LF around the relic area in the `sausage' cannot be attributed to accretion shocks, as these are expected to be generally symmetrical around the cluster outskirts, in the absence of a nearby obvious mass donor. ROSAT imaging \citep{1999A&A...349..389V} reveals that the `sausage' is possibly connected to another cluster located to the north-east via a filament. Despite many galaxies being expected to lie along this filament, we do not find any evidence of enhanced star-forming galaxy counts in its direction. We stress that despite the fact the `sausage' and the `toothbrush' clusters are in the Galactic plane and are affected by differential dust extinction across the FOV, we have corrected for effects arising from incompleteness.

We find that the morphologies and star-forming properties of the galaxies hosted by the `sausage' and `toothbrush' clusters are vastly dissimilar. Despite having matching masses, temperatures, orientations and radio morphologies, there is one crucial factor that might explain the striking difference between the two clusters. Simulations \citep{2011MNRAS.418..230V, 2012MNRAS.425L..76B} tailored to reproduce the radio morphologies of the two fields indicate that, while the `sausage' is a pure merger between two massive clusters, in the `toothbrush' an extra smaller sub-cluster is participating in the merger towards the end of the merging process. Moreover, the core passage in the `toothbrush' might have occurred $\sim2$~Gyr ago, while in the `sausage' only $\sim1$~Gyr ago. The `toothbrush' could be in a more advanced stage of relaxation, possibly close to virialisation (typical virialisation times for clusters are $\sim1$~Gyr), and the galaxy population evolved into gas-poor ellipticals. X-ray observations of the two clusters \citep{2013MNRAS.429.2617O, 2013MNRAS.433..812O} also reveal that the more massive subcluster is located to the north of the `sausage', in the proximity of the Mpc-wide relic, while for the `toothbrush' it is located towards the southern relic. H$\alpha$ is particularly sensitive to young, massive stars and it captures only recent episodes of star-formation. Shock compression is expected to be an `instantanous' process, exciting star formation momentarily as it passes through a galaxy. In order to detect a boost in the luminosity function, we need to observe a cluster at the adequate time when shock-induced star formation is still active within gas-rich galaxies. Since we are viewing the `toothbrush' cluster at a more evolved `time-slice', it is possible that there are fewer gas-rich galaxies for the shock to `light-up'.  
 
\section{CONCLUSIONS}\label{sec:conclusion}
In this paper we have analysed H$\alpha$/emission-line galaxies within vastly disturbed clusters hosting radio-identified shock fronts. As test cases, we studied the `sausage' and `toothbrush' clusters at redshift $\sim0.2$, which have been likely formed through major mergers in the plane of the sky. The merger gave rise to Mpc-wide coherent shock waves that produced spectacular examples of radio-relics. We used the H$\alpha$ recombination line, observing the clusters via the narrow-band technique with the Isaac Newton Telescope. We inspected individual H$\alpha$ morphologies and looked at global, statistical properties via H$\alpha$ luminosity functions.
\begin{itemize}
\item We surveyed a comoving volume of $3.371\times10^3$~Mpc$^3$ for the `sausage' cluster and $4.546\times10^3$~Mpc$^3$ for the `toothbrush'. We detected a total of 181 line emitters for the former and 141 for the latter, out of which $49$ and $30$ are expected to be H$\alpha$ emitters.

\item We produced field LFs encompassing the entire field-of-view of the two clusters and found good agreement with the COSMOS blank field LF derived by \citet{2008ApJS..175..128S} for a similar redshift. 

\item We discovered numerous extended galaxies around the radio relics in the `sausage' cluster. The normalisation of the H$\alpha$ luminosity function is an order of magnitude above that of luminosity functions derived for both local, relaxed clusters like Abell 1367 and Coma, and similar redshift, disturbed clusters like Abell 521. We speculate these extended H$\alpha$ haloes result from interactions with the travelling shock waves that are responsible for accelerating the radio-electrons. 

\item In the `sausage' cluster we uncover a clear boost ($9\sigma$) in the number counts of luminous H$\alpha$ emitters around the radio relics and speculate that the passage of the shock wave might have induced star-formation within the disks of the cluster galaxies. 

\item We do not find such bright emitters in the `toothbrush' cluster. We speculate that the difference between the emitter populations in the `sausage' and `toothbrush' clusters by examining their different merger history, particularly the time since core-passage. 
\end{itemize}

The results presented here show the potential of multi-wavelength studies of clusters hosting radio relics. Surveying a statistical sample of clusters containing radio relics would enable us to test the statistical properties of galaxies under the influence of travelling shock waves. Do shock fronts lead to a boost in the star formation activity, or is it the case only in the `sausage' cluster? Larger samples of clusters could reveal any dependence of shock-induced star-formation on the merging history of the clusters.

\section*{Acknowledgements}
We thank the referee for the useful comments which helped improved the clarity of the paper. We thank Jarle Brinchmann for useful discussions. The Isaac Newton Telescope is operated on the island of La Palma by the Isaac Newton Group in the Spanish Observatorio del Roque de los Muchachos of the Instituto de Astrof{\'i}sica de Canarias. This research has made use of the NASA/IPAC Extragalactic Database (NED) which is operated by the Jet Propulsion Laboratory, California Institute of Technology, under contract with the National Aeronautics and Space Administration. This research has made use of NASA's Astrophysics Data System. AS acknowledges financial support from NWO. DS is supported by a VENI fellowship. RJvW acknowledges support provided by NASA through the Einstein Postdoctoral grant number PF2-130104 awarded by the Chandra X-ray Center, which is operated by the Smithsonian Astrophysical Observatory for NASA under contract NAS8-03060.

\bibliographystyle{mn2e.bst}
\bibliography{Halpha_sausage_tooth}

\appendix

\section{H$\alpha$ fractions}
\begin{figure}
\begin{center}
\includegraphics[trim=0cm 0cm 0cm 0cm, width=0.495\textwidth]{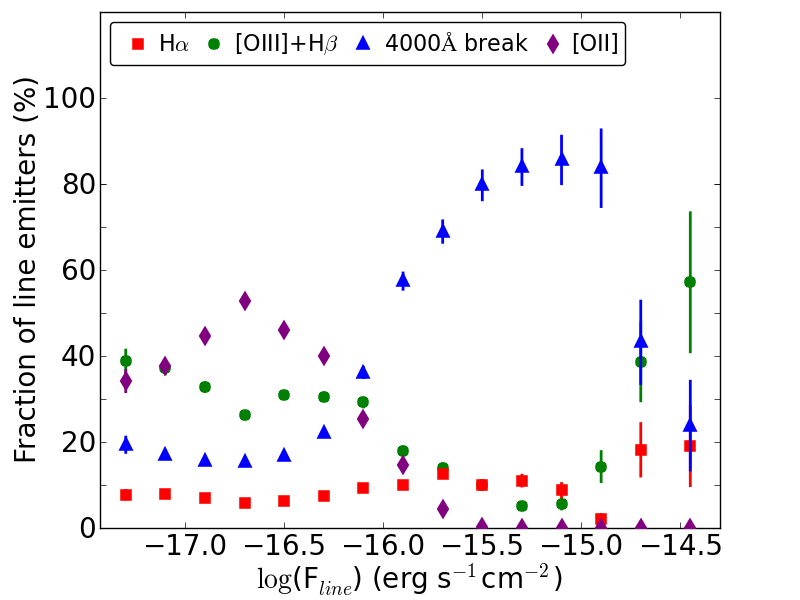}
\includegraphics[trim=0cm 0cm 0cm 0cm, width=0.495\textwidth]{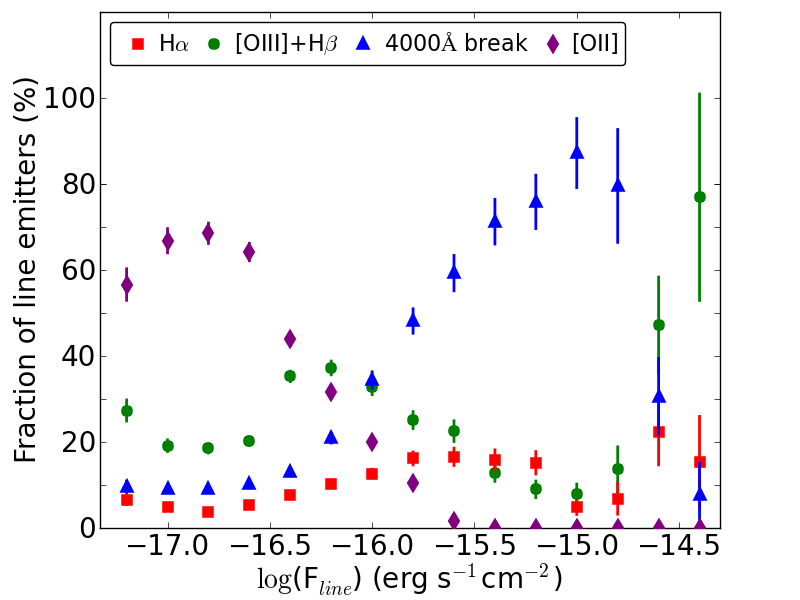}
\end{center}
\caption{H$\alpha$ and other emitters' fractions with respect to the total number of emitters detected by the narrow band filters, as a function of total line flux. \textit{Top}: `Sausage'-like data. \textit{Bottom}: `Toothbrush'-like catalogue. We used the COSMOS catalogue to simulate `sausage'-like and `toothbrush'-like catalogues. Error bars are overplotted, but in some cases they are smaller than the plotting symbol.}
\label{fig:fractions}
\end{figure}

\label{lastpage}
\nocite{*}
\end{document}